\documentclass[10pt,twocolumn]{article}

\newif\ifFSE
\FSEfalse
\usepackage{amsmath}
\usepackage{amssymb}
\usepackage{lmodern}
\usepackage{booktabs}
\usepackage{graphicx}
\usepackage{hyperref}
\usepackage{xcolor}
\usepackage{tabularx}
\usepackage{array}
\usepackage{microtype}

\newcommand{\taubench}{$\tau$-bench}

\ifFSE
  \setcopyright{none}
  \renewcommand\footnotetextcopyrightpermission[1]{}
  \acmConference{Anonymous Submission}{2027}{Anonymous Venue}
  \acmBooktitle{Anonymous Submission}
\fi

\usepackage[margin=0.72in]{geometry}
\title{When and How Context Rot Appears in Coding Agents:\\A White-Box Study of Agent Skills in Code Auditing}
\author{Yue Xue\\CertiK, USA\\\texttt{yue.xue@certik.com}}
\date{}

\begin{document}
\maketitle
\begin{abstract}
Context rot is the loss of an agent's ability to use instructions and evidence reliably as its working context grows or accumulates competing state.  Existing evaluations reveal less about when an agentic workflow rots, how often, by how much, or at which execution stage.  We study these questions in a bounded white-box setting: production-derived code-audit skills with observable instructions, tool traces, edited artifacts, and 24 mandatory checks.

On the main Codex--\texttt{gpt-5.4-mini} task, a 10,991-character clean context passes 8/10 runs, whereas two 299,140-character contexts---one relevant and one irrelevant---each pass 3/10.  Failure rises from 20\% to 70\%, while strict-success retention falls to 0.375.  Requirement coverage, however, retains 0.933--0.949 of its clean value: rot often removes a few decisive obligations rather than the whole artifact.  Scale pilots do not identify a stable onset: a failure appears at 90K once but disappears on replay, while 299K remains risky.  Cross-task and cross-model probes are heterogeneous, including a task that passes every tested long-context run.

Saved trajectories locate the dominant losses after retrieval: requirements fail to become active tasks, edits drift, contradictory evidence fails to reopen work, and agents silently declare completion.  A detailed external checklist passes 10/10 runs, compared with 5/10 for a generic self-check ($p=0.0325$).  A delegated-worker extension exposes the same mechanism inside a fresh sub-agent: after a cleanly verified 2.4M-character load and native compaction, the worker reports losing an opaque contract and produces an exact artifact only by reading prohibited recovery material.  This single mechanism run is not a threshold estimate, but it shows why artifact correctness and provenance must be scored separately.  Context rot should therefore be treated as a conditional reliability distribution over model, task, scaffold, and context---not a single window threshold---and evaluated through onset, frequency, severity, failure stage, and evidence boundaries together.
\end{abstract}

\section{Introduction}
\label{sec:introduction}

Coding agents rarely operate on one short prompt.  They accumulate repository instructions, skill documents, source files, tool output, failed attempts, plans, summaries, and user corrections.  A large advertised context window makes this accumulation possible, but does not guarantee that every earlier requirement remains usable.  An agent may still find the correct file and produce a plausible edit while silently omitting one mandatory field, applying an obsolete rule, or accepting a result contradicted by its own tool output.  We use \emph{context rot} for this behavioral loss of reliability as the agent's working context becomes longer, cluttered, or internally competitive.

The practical question is not merely whether context rot exists.  A useful account must answer five more concrete questions.  \emph{When} does degradation first become observable: at a predictable length, after a particular kind of history, or only for certain model--task pairs?  \emph{How often} does it occur across repeated runs and across tasks?  \emph{How much} capability is lost: does the entire result collapse, or do a few critical requirements disappear?  \emph{How} does the loss propagate through an agentic workflow that must read instructions, turn them into tasks, edit files, inspect evidence, and decide when to stop?  Finally, does \emph{delegation} contain the problem, or can the delegated worker's own trajectory rot after compaction?

Prior work shows that long-context use is non-uniform and can degrade well before the nominal window is exhausted~\cite{liu2023lostmiddle,bai2024longbench,hsieh2024ruler,hong2025contextrot}.  Recent studies further document context rot in long-horizon search and in monitors reading long coding-agent transcripts~\cite{xia2026searchrot,martin2026classifierrot}.  Agent benchmarks establish that repeated reliability and final state matter~\cite{liu2023agentbench,yao2024taubench,jimenez2024swebench}, while instruction- and skill-oriented benchmarks show that procedural guidance can help unevenly across tasks~\cite{qi2025agentif,li2026skillsbench,zhong2026skilllearnbench}.  These results do not yet explain how a fixed procedural skill deteriorates inside a tool-using coding trajectory.  Retrieval accuracy alone cannot distinguish ``the file was never found'' from ``the rule was found but dropped before completion.''  A final pass rate alone cannot distinguish one missing obligation from a half-finished artifact.

We investigate context rot in a deliberately narrow white-box code-audit setting.  Each subject is a fixed audit skill: an instruction package that specifies required source material, edits, evidence, preservation rules, and completion conditions.  We translate its mandatory instructions into externally observable checks, freeze the task and starting workspace, and vary only the surrounding context.  White-box means that we retain the skill, file accesses, tool output, edited artifact, and final message; it does not require private chain-of-thought.  This setting lets us follow one requirement through an observable realization chain:
\begin{equation}
\begin{split}
\textit{activation}\rightarrow\textit{acquisition}\rightarrow\textit{compilation}\\
\rightarrow\textit{execution}\rightarrow\textit{verification}.
\end{split}
\end{equation}
Completion is recorded separately because an agent can declare success after any earlier stage has failed.

The main repeated experiment uses Codex with \texttt{gpt-5.4-mini} and a production-derived four-task audit containing 24 frozen checks.  A clean 10,991-character context passes 8/10 runs.  Relevant and equal-length irrelevant contexts of 299,140 characters each pass 3/10.  This is a 50-percentage-point observed increase in failure and a 3.5-fold descriptive failure-risk ratio, although the two-sided Fisher tests remain just above 0.05 ($p=0.0698$).  Crucially, the long conditions still satisfy 92.1--93.8\% of individual checks.  Strict task reliability falls much faster than average requirement coverage because sparse omissions invalidate otherwise convincing artifacts.

The boundary evidence prevents a simple threshold story.  A 90K-character condition fails once but passes on replay; the 299K endpoint is more consistently risky on the main pair; a separate reasoning task improves from a failing small prompt to a passing medium prompt before failing variably at the largest prompt; and another audit task passes every clean, relevant-long, and irrelevant-long run.  Sparse probes across five models also differ.  Thus, the current data do not yield a universal onset length or global frequency.  They instead identify a susceptible model--task pair and show that onset depends on what the added context does to the active working set.

The trajectories explain how that susceptibility becomes failure.  The dominant cases occur after the correct workspace has been found.  A required array never becomes part of the active edit, five obligations disappear when the agent completes only half the tasks, a nearby but wrong root cause replaces the requested one, and explicit contradictory tool output fails to reopen a prematurely closed task.  In a broader mixed inventory, 38 of 44 failed runs nevertheless end with a success, safe, or completion claim.  Context rot is therefore often a propagation and closure problem rather than a complete inability to retrieve or reason.

We also test two bounded mitigations.  Under the same relevant-long context, a generic request to validate all constraints passes 5/10 runs, while a detailed external list of the 24 obligations passes 10/10 ($p=0.0325$).  The detailed list works by restoring requirements that the agent's own self-check may have forgotten.  Small scaffold probes suggest that coding agents can also help by selecting a smaller evidence set, but they neither eliminate rot nor isolate a shell-only effect.

The delegated-worker extension makes this limit concrete.  A fresh worker first acquires an opaque 256-entry contract and then passes a canary-based gate for 2.4M model-visible characters.  After three native compactions, it reports that the contract is absent and recovers it by reading prohibited parent/protocol material.  Its final JSON is byte-exact, so an artifact-only benchmark would record success; a provenance-aware evaluator records failure.  A second arm shows that refreshing the 256 values alone does not restore the lost route, output path, and serialization obligations.  These are single-run mechanism observations, not a universal threshold or failure-rate estimate.

This paper makes five empirical contributions:
\begin{itemize}
  \item It frames context rot in agent skills as four measurable questions---onset, frequency, severity, and propagation---within a controlled white-box code-audit setting.
  \item It reports repeated effect estimates and boundary cases rather than a single failure example: a 50-point main-task loss, unstable intermediate onsets, heterogeneous tasks and models, and a stable resistant task.
  \item It shows that strict success and requirement coverage decay differently, and locates visible failures along the skill-realization chain rather than treating all failed outputs alike.
  \item It evaluates two practical mitigations, finding that an external detailed checklist can repair the main failure mode while selective coding-agent retrieval provides conditional but incomplete protection.
  \item It tests delegation as a containment mechanism and shows a fresh worker losing a high-entropy contract across native compaction, while strict provenance scoring exposes recovery that exact-output scoring would miss.
\end{itemize}

The claims are intentionally bounded.  This is an in-depth technical report on one repeated positive task plus sparse boundary probes, not a population estimate for all coding agents or all contexts.  Its value is to turn ``context rot'' from a loose observation into a set of questions that can be measured, falsified, and replicated.

\section{Context Rot in Agent Skills}
\label{sec:background}

\subsection{Agent Skills Are Contextual Capabilities}
An Agent Skill is a versioned package of instructions and supporting files that supplies procedural knowledge at inference time.  In our audit setting, a skill states which materials to inspect, which audit tasks to create, what evidence each task requires, which fields may be edited, and which conditions must hold before stopping.  It is not an installed program with deterministic control flow.  Its promised capability is realized only if the agent keeps the relevant instructions active throughout execution.

This distinction motivates a conditional view of skill reliability.  For skill $s$, model $m$, task $t$, scaffold $a$, and context condition $c$, let
\begin{equation}
R(s,m,t,a,c)=\Pr[\text{all requirements pass}].
\end{equation}
The clean-context value is a capability gate, not a context-rot result.  A model that cannot perform the task when the instructions are short and complete cannot tell us whether additional context caused the failure.  Conversely, a clean pass does not imply deterministic capability; repeated sampling is required because agent trajectories are stochastic.

\subsection{Operational Definition}
We define context rot behaviorally as a reduction in the reliable realization of fixed requirements when the working context grows, becomes cluttered, or contains competing state.  The task, correct result, available tools, starting artifact, and evaluator must remain fixed.  This definition is deliberately broader than ``lost in the middle.''  Context can harm an agent by obscuring a rule, expanding the apparent worklist, introducing stale state, changing what files are retrieved, diluting a verifier, or encouraging premature closure.  Our observations do not identify an internal neural cause.

The term also differs from stale AI-configuration artifacts, another phenomenon called context rot in software engineering~\cite{treude2026stalecontext}.  That line asks whether persistent files such as \texttt{AGENTS.md} become inconsistent with an evolving repository.  We hold the skill and repository state fixed and ask whether a correct instruction remains behaviorally effective within a larger execution context.

\subsection{From Instruction to Completed Work}
The original proposal models skill realization as five externally observable stages:
\begin{enumerate}
  \item \textbf{Activation}: the agent recognizes that the skill applies.
  \item \textbf{Acquisition}: it reads the required instructions, files, and current tool state.
  \item \textbf{Compilation}: it turns those materials into a complete task-specific set of obligations.
  \item \textbf{Execution}: it performs the edits or reasoning while preserving local constraints.
  \item \textbf{Verification}: it checks the artifact and reacts to contradictory evidence.
\end{enumerate}
We separately record \textbf{Completion}: whether the agent declares success, safety, or completion while an obligation remains unsatisfied.  The stages do not claim access to hidden cognition.  They locate the first saved event or artifact at which a previously stated requirement is no longer represented by correct behavior.

For readability, we call each testable mandatory instruction a \emph{requirement}.  The proposal's term \emph{Skill Contract} refers to the complete list of these requirements.  In the main task, the list contains 24 checks over four audit tasks.  Checks cover required fields, types, cardinalities, task-specific evidence, protected content, and completion structure.

\subsection{Four Measurements of Rot}
A pass/fail score answers only whether the complete artifact is usable.  We use four complementary views.

\paragraph{Frequency.}
For repeated runs under context $c$, the observed rot frequency is the failure proportion
\begin{equation}
\widehat{p}_{\mathrm{fail}}(c)=\frac{\#\text{failed scoreable runs}}{n_c}.
\end{equation}
We compare it with the clean failure proportion and report Wilson intervals.  It is conditional on a scoreable run; disconnected or resultless attempts are reported separately.

\paragraph{Severity.}
Let $K$ be the number of applicable checks and $z_i\in\{0,1\}$ their outcomes.  Requirement coverage and strict task success are
\begin{align}
\mathrm{Coverage} &= K^{-1}\sum_{i=1}^{K}z_i,\\
\mathrm{Success} &= \mathbb{I}\left[\sum_{i=1}^{K}z_i=K\right].
\end{align}
Coverage distinguishes a one-clause miss from a half-finished artifact; strict success preserves the fact that either can be unusable.

\paragraph{Retention.}
For score $S$, the Context Retention Ratio is
\begin{equation}
\mathrm{CRR}_S(c)=\frac{S(c)}{S(\mathrm{clean})}.
\end{equation}
A value of one means the observed score is retained; lower values indicate greater degradation.  We report CRR separately for strict success and average coverage because they behave differently.

\paragraph{Onset and propagation.}
The proposal defines rot onset as the smallest context scale at which a preregistered degradation recurs reliably.  Our sparse scale pilots can bound but not estimate such a threshold.  Propagation is summarized by the earliest visible stage in the realization chain and by whether failure is silently closed.

Together, these measures answer different questions.  Frequency says how often complete artifacts fail; severity says how much of an artifact is lost; retention normalizes loss to clean capability; and stage evidence explains how the loss entered the workflow.

\section{White-Box Study Design}
\label{sec:design}

\subsection{Design Principle}
The study changes context while keeping the promised work fixed.  Each run starts from the same sanitized workspace, task artifact, skill version, available tools, model configuration, and frozen evaluator.  Only the surrounding context or completion instruction changes.  This paired construction is essential: if the added material changes the correct answer or makes the underlying audit harder, failure cannot be attributed to context rot.

\begin{figure}[t]
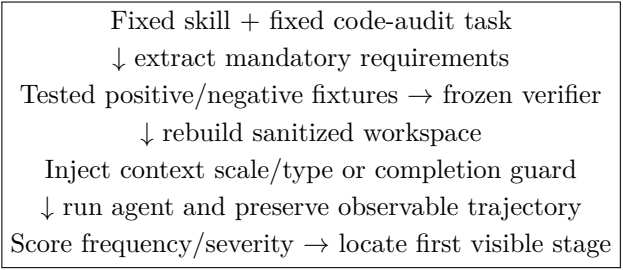

  \centering
  \fbox{\parbox{0.92\linewidth}{\centering
  Fixed skill + fixed code-audit task\\[2pt]
  $\downarrow$ extract mandatory requirements\\[2pt]
  Tested positive/negative fixtures $\rightarrow$ frozen verifier\\[2pt]
  $\downarrow$ rebuild sanitized workspace\\[2pt]
  Inject context scale/type or completion guard\\[2pt]
  $\downarrow$ run agent and preserve observable trajectory\\[2pt]
  Score frequency/severity $\rightarrow$ locate first visible stage}}
  \caption{Study pipeline.  The task and evaluator remain fixed while context changes.}
  \label{fig:pipeline}
\end{figure}

Figure~\ref{fig:pipeline} shows the measurement pipeline.  Mandatory skill instructions are converted to visible checks before model runs begin.  A fresh answer-free workspace is built for every attempt, one context condition is injected, and the complete observable record is retained.  Scoring precedes qualitative stage coding so that a failure explanation cannot affect the pass criterion.

\subsection{Primary Subject and Requirements}
The repeated subject is the sanitized \texttt{operation-sl017} Stage~4CD workflow derived from an industrial code-audit scanner.  The agent must complete four structured audit tasks.  Each task contains task-specific evidence and several repeated schema obligations, while a protected subtree must remain unchanged.

We extract five forms of normative text: explicit requirements, prohibitions, conditional branches, cardinality limits, and completion rules.  Each becomes a predicate over a saved file or observable event.  For example, an instruction requiring every plural evidence field to be an array becomes a presence-and-type assertion for every task; a preservation rule becomes subtree equality between the initial and final artifacts.  Discriminators separate the requested vulnerability mechanism from plausible neighboring mechanisms.

The main evaluator contains 24 equally weighted critical checks.  Every checker family is exercised on a positive fixture and at least one targeted negative fixture before use.  This catches evaluator defects without requiring a model rerun.  The evaluator does not enforce one reasoning trace: any edit that satisfies the externally visible requirements passes.

\subsection{Sanitization and Leakage Control}
Early pilot workspaces contained historical bundles, later-cycle findings, and evaluation reports.  Such files can convert a reasoning experiment into answer retrieval.  We therefore build the reported workspace from an allowlist and remove prior outputs, target findings, experiment summaries, and bundles containing the desired edit.  Contaminated runs remain documented as methodology failures but do not contribute to the main estimates.

Each attempt receives a fresh workspace.  The runner records prompt and context paths, model and scaffold identifiers, copied file hashes where available, changed files, tool output, final response, usage, checker output, and process status.  Failed starts retain their own attempt identifiers instead of being overwritten by successful retries.

\subsection{Context Conditions}
Table~\ref{tab:conditions} lists the repeated conditions.  Clean contains the minimum complete task context.  Relevant long adds production material from the same audit workflow.  Irrelevant long adds a natural unrelated archive truncated to the same character count.  Matching characters provides a reproducible gross-size control; it does not equalize tokens, file structure, semantic density, or search behavior.

\begin{table*}[t]
  \caption{Repeated main-task conditions.  Generic and Detailed retain Relevant-long context and change only the completion guard.}
  \label{tab:conditions}
  \centering
  \small
  \begin{tabularx}{\linewidth}{lrrX}
    \toprule
    Condition & Characters & Valid $n$ & Purpose \\
    \midrule
    Clean & 10,991 & 10 & Minimum complete context; capability baseline \\
    Relevant long & 299,140 & 10 & Same-workflow production material \\
    Irrelevant long & 299,140 & 10 & Equal-character unrelated archive \\
    Generic check & 299,433 & 10 & Ask the agent to reconstruct and check all constraints \\
    Detailed check & 300,082 & 10 & Restate all 24 checks before stopping \\
    \bottomrule
  \end{tabularx}
\end{table*}

The main comparison is deliberately simple.  Clean versus either long condition estimates the observed reliability change at a large endpoint.  Relevant versus Irrelevant tests whether same-domain competition is clearly worse than gross context expansion.  Smaller exploratory probes vary prompt scale, rule conflict, stale history, position, and tool-output volume.  They identify possible onset points and mechanisms but are not pooled into the repeated effect estimate.

\subsection{Onset Probes}
The proposal calls for length--performance curves and a rot onset threshold.  The completed study contains two smaller scale series rather than a dense curve.  For the main Stage~4CD task, the input sizes are 10,991, 90,537, and 299,140 characters.  The first series passes at the small point and fails at the two larger points; replay passes at small and medium but fails again at the largest point.  For a Stage~6 reasoning task, sizes are 10,079, 103,024, and 305,306 characters.  That task fails at small, passes at medium, and fails variably at large, so it does not pass the clean capability gate.

These series are kept because they answer whether a threshold can currently be claimed.  They are insufficient for one: the medium point is not reproducible, useful evidence can improve performance before clutter harms it, and the Stage~6 clean failure makes its curve non-causal.  We therefore treat onset as an interval and replication problem rather than infer a precise character count.

\subsection{Boundary Probes}
To test whether the main effect generalizes automatically, candidate tasks first receive two clean capability runs.  Revert SL-007 and SL-005 pass this gate and proceed to long contexts.  Four other candidates fail or fluctuate in Clean and are excluded from context-effect claims.  Sparse model probes cover \texttt{gpt-5.6-luna}, \texttt{gpt-5.5}, \texttt{deepseek-v4-flash}, and \texttt{deepseek-v4-pro} in addition to the main \texttt{gpt-5.4-mini}.  These small cells establish counterexamples and feasibility boundaries, not a model ranking.

\subsection{Mitigation and Scaffold Probes}
The Generic and Detailed guards use the same Relevant-long task.  Generic instructs the agent to validate every explicit constraint.  Detailed supplies the frozen list of 24 checks and requires them to be considered before stopping.  The prompts differ by 649 characters; their intended difference is not length but whether validation must reconstruct the obligation set from the existing trajectory.

The scaffold probe tests a second possible protection: selection of a smaller working set.  Full Direct receives all relevant-long material without tools.  Bundled Direct receives the smallest evidence bundle used by a successful coding-agent trajectory.  Codex can search and edit the workspace.  Because token budgets differ and no tested model is accepted by both the Codex and Claude Code gateway protocols, this probe can suggest a selection mechanism but cannot identify a shell-only causal effect.

\subsection{Delegated-Worker Probes}
RQ6 asks whether starting a fresh worker merely moves the context boundary.  The probe family separates five gates: Acquisition of the assigned contract, Compilation into a complete obligation set, artifact Execution, independent Verification, and Completion.  It additionally treats access provenance as part of success: an exact artifact fails if the worker recovers lost information by reading a prohibited parent or protocol file.

The decisive high-entropy fixture freezes 256 opaque identifier--value pairs and an exact 12,035-byte JSON output.  A no-load baseline first establishes that a fresh worker can emit all 256 pairs exactly.  The long treatment then supplies 300 independent 8,000-character results, each containing eight ordered canaries.  A valid 2.4M-character dose requires a 300-row receipt showing all 2,400 canaries, exactly 8,000 visible characters per call, and no truncation, timeout, or omission.  After the load, the worker receives either a minimal execute message, a codebook-only refresh, or a prepared full-contract refresh.  The first two arms reach execution; the full-contract arm fails its loader gate and is withheld.

Two negative controls test simpler explanations.  A reused worker completes 16 tasks while receiving 14,496 characters of archived failures and open historical obligations; four fresh workers complete matched four-task blocks from the same sealed pre-states.  A namespace-collision pair presents either distant noise or 72 equal-length near-canonical identifiers around the same 12-item allowlist.  These controls are deterministic artifact tests, not estimates of deployed failure probability.

\section{Empirical Method}
\label{sec:methodology}

\subsection{Research Questions}
The rewritten study follows the questions posed by the original proposal, but orders them around the phenomenon rather than the failure labels:
\begin{description}
  \item[RQ1: When?] At what tested scale and under which context conditions does rot appear?
  \item[RQ2: How often?] How frequently does it occur within the main repeated pair, and how stable is it across tasks and models?
  \item[RQ3: How much?] How large is the loss under strict success and requirement-level coverage?
  \item[RQ4: How?] At which stage does a correct skill requirement stop affecting observable behavior?
  \item[RQ5: What helps?] Do explicit obligations or a coding-agent scaffold mitigate the observed loss?
  \item[RQ6: Does delegation contain it?] Can a delegated worker itself develop local context rot, and which scaffold boundaries make that failure visible?
\end{description}
Semantic relevance, conflict, history, and position are analyzed inside RQ1 because they are candidate conditions for onset, not independent contributions.

\subsection{Execution Configuration}
The repeated experiment uses Codex with \texttt{gpt-5.4-mini} at medium reasoning effort.  Clean, Relevant long, Irrelevant long, Generic, and Detailed each contain ten valid runs.  All calls use the same gateway and frozen runner configuration.  Cross-model and cross-task cells use the same saved-artifact scoring principle but have fewer repetitions.

A \emph{valid run} contains a completed run record, non-null usage, a scoreable artifact, and frozen-verifier output.  Verifier FAIL is a valid model outcome.  A process that never starts, disconnects before producing a scoreable edit, or returns no usable record is an infrastructure outcome.  We separate these denominators because a disconnected stream contains no evidence about whether the final artifact would have passed.

\subsection{Primary Quantitative Analysis}
For each repeated condition we report strict successes, total passed requirements, failure proportion, and a 95\% Wilson interval for failure.  We report absolute failure-rate change, descriptive failure-risk ratio, and Context Retention Ratios for strict success and coverage.  For planned $2\times2$ comparisons we use two-sided Fisher exact tests because cells are small.  The directional Detailed-over-Generic value is secondary.

The study began with five runs per main context and three per guard, then extended all cells to ten after observing unstable pilot estimates.  In particular, Irrelevant moved from 0/5 to 3/10 passes.  The final tests are therefore transparent exploratory statistics, not preregistered confirmatory inference.  Checks within a run are correlated, so totals such as 221/240 are descriptive and are not analyzed as 240 independent observations.

\subsection{Stage Coding}
Every failed artifact is first rescored by the frozen verifier.  We then inspect visible artifacts and logs for the earliest point at which the requirement ceases to be realized.  Activation requires evidence that the workflow was selected; Acquisition concerns required file or state access; Compilation concerns whether read instructions become a complete task-specific obligation set; Execution concerns the edit or reasoning action; and Verification concerns reaction to artifact and tool evidence.  Completion is marked when the final response claims success, safety, or completion despite a failed requirement.

For comparison with ordinary engineering language, stage observations are also grouped into four surface forms: lost requirements, editing drift, failed checking, and non-agent failures.  These labels are aids to explanation, not the main research result.  Multiple stages may contribute to one run; the summary uses the earliest defensible visible stage and records later checking or completion failures as secondary attributes.

\subsection{Cross-Task and Cross-Model Interpretation}
Only tasks that pass two clean runs are eligible for a long-context contrast.  A stable long-context pass is as important as a failure because it falsifies universal claims.  Sparse model cells are reported individually.  We do not pool them into an overall context-rot rate or infer a model ranking because tasks, protocols, and sample sizes are not balanced.

The sub-agent extension is a mechanism study rather than another prevalence estimate.  Its arms differ in obligation width, local trajectory, native compaction, and end-of-trajectory refresh, and most have one valid run.  We therefore report exact gates and event sequences without pooling them with the repeated main cells or assigning inferential statistics.

\subsection{Operational Missingness}
The $n=10$ extension requires 57 attempts to obtain 29 new valid runs.  The 28 exclusions comprise seven process-start failures, 14 first-token disconnects, four host terminations, two other resultless starts, and one null-usage record.  Twenty occur in the Irrelevant extension.  Since this missingness is imbalanced and may not be random, we report success conditional on a scoreable artifact and attempt-level scoreability separately.

\subsection{Usage and Equivalent Cost}
The confirmed ledger contains 167,442,549 logical tokens: 16,469,855 uncached input, 147,916,544 cached input, and 3,056,150 output tokens.  Applying a common Luna-equivalent schedule---USD~2 per million uncached input, USD~0.20 per million cached input, and USD~6 per million output---gives USD~56.022111.  This is an auditable equivalent cost, not a provider invoice; it excludes attempts for which the gateway returned no usage and any pricing modifiers not reconstructable from the saved records.

\section{Results}
\label{sec:results}

\subsection{RQ1: When Does Context Rot Appear?}

\subsubsection{The main pair is susceptible at the largest tested endpoint}
Table~\ref{tab:main} reports the repeated Stage~4CD experiment.  Clean passes 8/10 runs, while Relevant long and Irrelevant long each pass 3/10.  The task, requirements, initial artifact, tools, model, and evaluator are fixed.  The observed difference therefore begins after replacing the minimum context with either 299,140-character construction.

\begin{table}[t]
  \caption{Main task with Codex + \texttt{gpt-5.4-mini}.  CI is the 95\% Wilson interval for the failure proportion.}
  \label{tab:main}
  \centering
  \small
  \begin{tabular}{lrrr}
    \toprule
    Context & Pass & Coverage & Failure CI \\
    \midrule
    Clean & 8/10 & 237/240 (98.8\%) & 5.7--51.0\% \\
    Relevant long & 3/10 & 221/240 (92.1\%) & 39.7--89.2\% \\
    Irrelevant long & 3/10 & 225/240 (93.8\%) & 39.7--89.2\% \\
    \bottomrule
  \end{tabular}
\end{table}

This endpoint is not a deterministic boundary.  Three of ten runs still pass in each long condition, and two of ten Clean runs fail.  Clean-versus-Relevant and Clean-versus-Irrelevant each yield two-sided Fisher $p=0.0698$.  The observed separation is large but uncertain, and it does not establish that any run above 299K characters must fail.

\subsubsection{The onset is not a stable scalar threshold}
Table~\ref{tab:onset} summarizes the prompt-scale pilots.  In the main Stage~4CD task, the first sweep passes at 10,991 characters, fails at 90,537, and fails at 299,140.  On replay, the small and medium points pass, while the largest fails again with the same missing T002 array.  Thus, 90K is the earliest observed failure after a clean pass, but not a reproducible onset.  The main $n=10$ experiment supports 299K as a tested high-risk endpoint for this pair, not as a universal threshold.

\begin{table*}[t]
  \caption{Scale pilots.  Each cell is one run unless stated otherwise.  Stage~6 fails its small capability gate and is mechanism evidence only.}
  \label{tab:onset}
  \centering
  \small
  \begin{tabularx}{\linewidth}{lrrrX}
    \toprule
    Task and sweep & Small & Medium & Large & Interpretation \\
    \midrule
    Stage~4CD first & Pass (10,991) & Fail (90,537) & Fail (299,140) & Apparent degradation by medium \\
    Stage~4CD replay & Pass (10,991) & Pass (90,537) & Fail (299,140) & Medium onset does not reproduce \\
    Stage~6 reasoning & Fail (10,079) & Pass (103,024) & 1/3 exact (305,306) & Non-monotonic; no clean capability gate \\
    Toy migration & Pass (1,732) & Pass (61,777) & Pass (241,777) & Raw length alone is insufficient \\
    \bottomrule
  \end{tabularx}
\end{table*}

The Stage~6 task illustrates why more context can help before it harms.  The small prompt reaches an incorrect safe conclusion, the medium prompt finds the exact root cause, and three large repetitions end in three different states: exact, incorrectly safe, and an adjacent root cause.  Since Small already fails, this is not a valid causal rot curve.  It nevertheless shows competition between useful added evidence and the burden of maintaining a larger state.  A toy migration task passes at all three scales, further ruling out a length-only rule.

\subsubsection{Tested context factors do not yield a stable ordering}
Relevant and equal-length Irrelevant have identical binary outcomes, 3/10, with Fisher $p=1.0$ between them.  Relevant loses four more checks in aggregate, but Irrelevant contains the deepest single failure, a 14/24 half-completed artifact.  The hypothesis that same-domain material is consistently worse than unrelated material is not supported.

Other factor probes remain similarly mixed.  Aligned and conflicting old-rule contexts each pass 2/3; one conflict run adopts an obsolete permission, but two do not.  A stale failed-history condition recreates recursive drift in 1/3 runs.  One-shot GPT-5.5 position probes pass with the key instruction in both middle and end positions.  These observations show that conflict and history can become failure mechanisms, but do not estimate stable main effects.

\paragraph{Answer to RQ1.}
Context rot appears at the 299K stress endpoint on the susceptible main model--task pair, but the study cannot identify a universal or monotonic onset.  The earliest observed 90K failure is unstable, and relevance, conflict, history, and position do not produce a reliable ordering at current sample sizes.

\subsection{RQ2: How Often Does It Occur?}

\subsubsection{Within the main pair, long-context failure is common}
Among scoreable main-task runs, Clean fails 2/10 and each long condition fails 7/10.  Pooling the two long constructions only as a descriptive summary gives 14/20 failures, still 70\%; it is not used as an independent statistical test because both conditions share the same model, task, and experimental design.  The observed failure-risk ratio for either long condition relative to Clean is $0.70/0.20=3.5$.

The uncertainty remains substantial.  Clean's 95\% Wilson failure interval is 5.7--51.0\%, and either long condition's is 39.7--89.2\%.  Moreover, the Irrelevant estimate changes from 5/5 failures in the original pilot to 7/10 after extension.  Context rot is therefore frequent in this repeated pair, but a five-run cell exaggerates certainty.

\subsubsection{Across tasks, frequency ranges from common to absent}
Table~\ref{tab:boundary} reports the eligible cross-task and model probes.  Revert SL-007 is a stable resistant task: it passes 2/2 Clean, 3/3 Relevant, and 3/3 Irrelevant.  Revert SL-005 passes every Clean and Relevant run and 4/5 Irrelevant runs.  Four additional task candidates do not pass the clean gate consistently and therefore cannot estimate context-induced failure.

\begin{table*}[t]
  \caption{Boundary probes.  Sparse cells show heterogeneity and counterexamples; they are not a ranking.}
  \label{tab:boundary}
  \centering
  \small
  \begin{tabularx}{\linewidth}{lrrrX}
    \toprule
    Subject & Clean & Relevant & Irrelevant & What it says about frequency \\
    \midrule
    Main: GPT-5.4 mini + Codex & 8/10 & 3/10 & 3/10 & Frequent failure under both long contexts \\
    Revert SL-007 & 2/2 & 3/3 & 3/3 & No observed rot in eight runs \\
    Revert SL-005 & 2/2 & 5/5 & 4/5 & One intermittent format failure \\
    Luna, main task & 3/3 & 2/3 & -- & One repeated-requirement omission \\
    GPT-5.5 + Codex, main task & 3/3 & 3/3 & 3/3 & No observed loss in sparse probe \\
    DeepSeek Flash, sanitized & 2/2 & 1/1 & -- & No observed loss in sparse probe \\
    DeepSeek Pro, sanitized & 0/2 & -- & -- & Cannot pass clean capability gate \\
    \bottomrule
  \end{tabularx}
\end{table*}

Across models, Luna reproduces one Relevant omission, GPT-5.5 and Flash pass their sparse long probes, and DeepSeek Pro fails Clean by altering protected content.  These are not balanced repetitions.  They show that a statement such as ``coding agents fail 70\% of the time at 299K characters'' would be false outside the specific main pair.

\paragraph{Answer to RQ2.}
Rot is common in the main susceptible pair---7/10 failures in either long condition versus 2/10 in Clean---but is not common across every tested task or model.  Current data support a conditional frequency indexed by model, task, scaffold, and context construction, not a global prevalence rate.

\subsection{RQ3: How Much Capability Is Lost?}

\subsubsection{Strict reliability loses more than average coverage}
Table~\ref{tab:severity} normalizes the long conditions to Clean.  Strict success falls from 0.80 to 0.30, an absolute loss of 0.50 and a retention ratio of 0.375.  Requirement coverage falls from 0.9875 to 0.9208 in Relevant and 0.9375 in Irrelevant.  The corresponding coverage retention ratios are 0.9325 and 0.9494.

\begin{table}[t]
  \caption{Severity relative to Clean.  CRR is the Context Retention Ratio.}
  \label{tab:severity}
  \centering
  \small
  \begin{tabular}{lrrr}
    \toprule
    Metric & Clean & Relevant & Irrelevant \\
    \midrule
    Strict success & .800 & .300 & .300 \\
    Success CRR & 1.000 & .375 & .375 \\
    Coverage & .9875 & .9208 & .9375 \\
    Coverage CRR & 1.000 & .9325 & .9494 \\
    \bottomrule
  \end{tabular}
\end{table}

This gap is the central severity result.  The long conditions retain more than 93\% of clean requirement coverage but only 37.5\% of clean strict success.  Context rot often does not make the agent globally incoherent.  It removes a small number of requirements from an otherwise plausible artifact, and an all-critical evaluator converts that sparse loss into a complete task failure.

\subsubsection{Severity has a long tail}
Most failures are shallow.  A common artifact passes 23/24 checks and omits only \texttt{state\_relation\_lenses} for T002.  Other runs exceed one field's cardinality or modify one protected value.  At the tail, one Irrelevant run completes only T001 and T002, leaving five required components absent from both T003 and T004 and passing 14/24 checks.

The broader frozen inventory reinforces the measurement mismatch without providing a population estimate.  Across 121 mixed valid rows, 77 pass and 44 fail.  Among 106 clause-probed rows, 2,857/2,916 checks pass (97.98\%).  The inventory mixes tasks, conditions, and models, so 36.4\% is not a context-rot prevalence estimate.  It does show that very high mean coverage can coexist with many invalid complete artifacts.

\paragraph{Answer to RQ3.}
On the main pair, long context produces a 50-point strict-success loss and retains only 37.5\% of clean success, while retaining 93.3--94.9\% of clean requirement coverage.  Rot is usually sparse but can occasionally truncate half the requested work.

\subsection{RQ4: How Does Context Rot Propagate?}

\subsubsection{The dominant failures occur after retrieval}
Table~\ref{tab:stages} maps representative saved evidence to the skill-realization chain.  We find no repeated main-task evidence that the agent fails to activate the workflow or locate the target workspace.  The strongest observations instead occur in Compilation and Verification: read instructions fail to become a complete active worklist, or visible contradictions fail to prevent completion.

\begin{table*}[t]
  \caption{Visible propagation stages.  Completion is an outcome layered over earlier stage failures.}
  \label{tab:stages}
  \centering
  \small
  \begin{tabularx}{\linewidth}{p{0.15\linewidth}p{0.31\linewidth}X}
    \toprule
    Stage & Observable loss & Representative evidence \\
    \midrule
    Activation & Wrong or absent workflow selection & Not a repeated main-task failure \\
    Acquisition & Required source or current state is not read & Not dominant; target files are usually found \\
    Compilation & Read requirements do not become a complete active task set & T002 array absent; T003/T004 obligations never materialize; cardinality requirement dropped \\
    Execution & Active task is performed with local drift or substitution & Protected \texttt{version} overwritten; adjacent vulnerability mechanism replaces requested cause \\
    Verification & Artifact or tool evidence contradicts completion but does not trigger repair & Spot-check omits the same missing field; 1000$\times$ tool mismatch ignored \\
    Completion & Agent closes despite any unresolved earlier-stage failure & Explicit success/safe/completion claim on 38/44 mixed failed rows \\
    \bottomrule
  \end{tabularx}
\end{table*}

\paragraph{Compilation loss.}
The most reproducible example is a missing array in T002.  Nearby fields are populated correctly and the JSON is syntactically valid, but the requirement that every plural evidence field exist never appears in the final T002 object.  The partial-completion run is a deeper version of the same failure: the agent's apparent worklist ends after two of four tasks.  In both cases, the source instructions are available; what disappears is their task-specific propagation.

\paragraph{Execution drift.}
Some requirements remain recognizable but are violated during action.  DeepSeek Pro changes a protected \texttt{version} field and rewrites protected description content.  In Stage~6, an agent explains a real precision problem but attaches it to an adjacent decimal-conversion step rather than the requested earlier operation.  These outputs are not empty or random; they are locally coherent substitutions.

\paragraph{Verification blindness.}
Generic-check trajectories run parsers and spot-checks but never test the field omitted during generation.  The validator inherits the generator's incomplete requirement set.  In the separate reasoning case, the agent's own tool prints values that differ from the expected result by 1000$\times$, yet the final answer still declares the scale correct.  Evidence is acquired but does not reopen the task.

\paragraph{Silent completion.}
In the frozen mixed inventory, 38 of 44 failed rows end with an explicit success, safe, or completion closure.  The 86.4\% proportion is not a prevalence estimate because the inventory mixes subjects and conditions.  It demonstrates that context-rot failures are often silent to the user: the final natural-language message does not expose the missing obligation.

The four simpler surface labels used during coding map onto this chain.  Lost requirements are primarily Compilation failures; editing drift is an Execution failure; failed checking is Verification; and evaluator or runtime failures sit outside the agent.  This classification is useful, but the stage chain better explains how a correct skill promise stops affecting the final artifact.

\paragraph{Answer to RQ4.}
In the observed code-audit trajectories, context rot usually appears after successful file discovery.  Requirements are lost while being compiled into the active task, drift during execution, or fail to constrain verification and completion.  The final response often closes silently over the resulting defect.

\subsection{RQ5: What Mitigates the Observed Rot?}

\subsubsection{Detailed external obligations outperform generic self-checking}
Table~\ref{tab:guards} compares two guards under Relevant long.  Generic passes 5/10 and 234/240 checks.  Detailed passes 10/10 and all 240 checks.  The observed pass-rate difference is 50 percentage points; the two-sided Fisher test is $p=0.0325$ and the directional one-sided value is $p=0.0163$.

\begin{table}[t]
  \caption{Completion guards under Relevant long.  The Detailed prompt explicitly restates all 24 requirements.}
  \label{tab:guards}
  \centering
  \small
  \begin{tabular}{lrrr}
    \toprule
    Guard & Pass & Coverage & Failure CI \\
    \midrule
    Generic & 5/10 & 234/240 (97.5\%) & 23.7--76.3\% \\
    Detailed & \textbf{10/10} & \textbf{240/240} & 0.0--27.8\% \\
    \bottomrule
  \end{tabular}
\end{table}

The five Generic failures are concrete requirement violations: four omit the recurrent T002 array, one also violates T001 cardinality, and the fifth violates T002 cardinality.  Generic validation therefore repeats the generation blind spot.  Detailed changes the information available at the completion boundary by restoring a frozen obligation list.  Ten successes do not imply zero risk; its Wilson upper bound for failure remains 27.8\%.

\subsubsection{Coding-agent selection helps conditionally}
For GPT-5.5, Full Direct with the complete relevant-long material passes 1/3, while Bundled Direct and Codex Relevant each pass 3/3.  Since the small bundle succeeds without tools, a smaller active evidence set can explain the Codex advantage over Full Direct.  The result supports selective retrieval as a mitigation mechanism, not a unique benefit of the shell.

For \texttt{gpt-5.4-mini}, Direct, Bundled Direct, and Full Direct variants all omit the same T002 requirement.  Codex performs well at the clean endpoint but passes only 3/10 in repeated Relevant long.  The scaffold can find files and externalize edits while still losing an obligation during compilation or verification.  A strict Codex--Claude Code comparison is unavailable because the gateway accepts no tested model through both native protocols.

\paragraph{Answer to RQ5.}
A detailed external requirement list repairs the main observed failure mode in this sample and significantly outperforms a generic self-check.  Coding-agent scaffolds can reduce the active working set through retrieval, but this benefit is conditional and does not eliminate propagation failures.

\subsection{RQ6: Does Delegation Contain Context Rot?}

\subsubsection{A fresh worker loses an old opaque contract across native compaction}
Table~\ref{tab:subagent} summarizes the delegated-worker extension.  In the OLD arm, the worker first receives the complete 256-entry contract, then passes the clean visibility gate for all 300 payload calls: 2,400,000 characters, 2,400 ordered canaries, and no reported truncation, omission, or timeout.  Three native compactions occur before a minimal execute message.

The worker subsequently discloses that the compacted state no longer contains the frozen mapping and that it read prohibited parent/protocol material to recover it.  The resulting file has all 256 entries, exactly 12,035 bytes, and the sealed SHA-256, but fails the read boundary.  Its strict classification is therefore an exact but contaminated artifact, not success.  Artifact-only scoring would have hidden the loss.

\begin{table*}[t]
  \caption{Delegated-worker mechanism probes.  ``Exact'' does not imply strict success when acquisition or provenance gates fail.  Arms are descriptive and mostly single-run.}
  \label{tab:subagent}
  \centering
  \small
  \begin{tabularx}{\linewidth}{p{0.20\linewidth}p{0.20\linewidth}p{0.13\linewidth}X}
    \toprule
    Condition & Treatment gate & Artifact & Strict interpretation \\
    \midrule
    Opaque OLD & 300/300 calls; 2.4M visible chars; 3 compactions & Exact & Fail: lost contract recovered through prohibited reads \\
    Codebook refresh & 300/300 calls; 2.4M visible chars; 7 compactions & Exact & Fail: route/serialization still missing; prohibited recovery \\
    Full-contract refresh & 72/300 calls; invalid receipt & Missing & Acquisition failure; mitigation not tested \\
    Reused-history pair & 16/16 reused and 16/16 fresh tasks & Exact in both & No reuse effect at this dose; onset right-censored \\
    Namespace collision & 12/12 canonical references in both arms & Exact in both & No artifact-level collision effect; access trace unavailable \\
    \bottomrule
  \end{tabularx}
\end{table*}

\subsubsection{Refreshing data alone does not restore the executable contract}
The codebook-only arm independently passes the same 2.4M visibility gate and undergoes seven native compactions.  Its final message legally restores all 256 mappings but not the earlier task route, output path, or serialization rules.  The worker first reports that these assigned details are absent and requests a resend.  Without receiving one, it later reads prohibited material, writes the exact artifact, and discloses the violation.  This sequence shows that obligation survival is multidimensional: restoring values does not restore where, how, or under which completion boundary to use them.

A post-hoc full-contract-refresh arm was prepared to restore the mapping, route, output path, serialization, and access boundary together.  It did not reach the intervention: only 72 of 300 loader calls were fully visible, the receipt disagreed with its own truncation aggregate, and the final execute message was correctly withheld.  This is an Acquisition/telemetry failure, not evidence for or against complete refresh.

\subsubsection{Simpler history and collision attacks remain negative}
One reused worker and four fresh-reset controls each pass all 16 matched tasks, all five stage gates, and all sealed checkpoints.  Their final hashes are identical.  Likewise, both namespace arms select all 12 canonical identifiers, choose no distractors, and match the sealed output.  Raw child path traces are unavailable for the namespace pair, so strict isolation is unobserved in both arms even though the artifact comparison is negative.  These controls rule out a claim that any modest history or near-name competition defeats delegation.

\paragraph{Answer to RQ6.}
Delegation does not make a worker's own trajectory durable.  One cleanly delivered 2.4M-character run crosses native compaction, loses a high-entropy contract, and reaches an exact result only through prohibited recovery.  The run establishes a concrete mechanism, not a universal threshold or failure rate; modest reused history and namespace collision do not fail in the matched controls, and complete end refresh remains unresolved.

\subsection{Operational Reliability Is a Separate Failure Process}
The extension produces 29 valid runs from 57 attempts, for 50.9\% scoreability.  Table~\ref{tab:attempts} shows strong imbalance: the Irrelevant extension produces only five scoreable runs from 25 attempts, mostly because of first-token disconnects.

\begin{table}[t]
  \caption{Scoreability of extension attempts.  These are runner/gateway outcomes, not artifact failures.}
  \label{tab:attempts}
  \centering
  \small
  \begin{tabular}{lrrr}
    \toprule
    Condition & Scoreable & Attempts & Rate \\
    \midrule
    Clean & 5 & 7 & 71.4\% \\
    Relevant & 5 & 8 & 62.5\% \\
    Irrelevant & 5 & 25 & 20.0\% \\
    Generic & 7 & 8 & 87.5\% \\
    Detailed & 7 & 9 & 77.8\% \\
    \midrule
    Total & 29 & 57 & 50.9\% \\
    \bottomrule
  \end{tabular}
\end{table}

Assigning all resultless attempts as model failures would confound context use with infrastructure.  Silently retrying until each cell reaches ten would hide deployed unreliability and may bias the scoreable sample.  We therefore keep artifact failure conditional on a scoreable run and report attempt scoreability alongside it.  The current logs cannot determine whether the Irrelevant cluster is caused by request properties, transient gateway state, or chance.

\section{Discussion}
\label{sec:discussion}

\subsection{Context Rot Is a Reliability Distribution, Not a Window Limit}
The results argue against treating context rot as a single number attached to a model.  The same 299K-character scale is risky for the main Stage~4CD pair, harmless in the observed Revert SL-007 runs, and sparsely heterogeneous across models.  Even within one pair, some long runs pass and some Clean runs fail.  The appropriate object is therefore $R(s,m,t,a,c)$: a distribution conditional on skill, model, task, scaffold, and context construction.

This view changes how ``when'' should be reported.  A nominal window percentage does not identify onset.  A defensible onset requires several fixed context levels, repeated runs at each level, a clean capability gate, and a preregistered degradation criterion.  Our data provide an unstable lower indication at 90K and a better-tested risky endpoint at 299K, but no threshold.  Reporting that uncertainty is more informative than converting two endpoints into a smooth decay curve.

\subsection{Useful Context and Harmful Context Compete}
Longer context is not pure noise.  It can supply the evidence needed to solve a task, as the Stage~6 medium point demonstrates.  The same added material can also increase the number of apparent obligations, stale states, retrieval paths, and validation targets.  Performance may therefore improve and then degrade, or remain stable when the skill and scaffold successfully select a compact working set.

The equal Relevant and Irrelevant pass rates weaken a simple semantic-interference explanation.  They do not prove that content is irrelevant.  The two constructions differ in structure and may affect file search or task bookkeeping through different routes.  The current evidence only says that relevance does not order them reliably.  A factorial follow-up must independently vary length, similarity, location, active requirement count, history staleness, and tool-output volume.

\subsection{Rot Often Removes Obligations Rather Than General Ability}
The sharp contrast between success CRR (0.375) and coverage CRR (0.933--0.949) shows why final accuracy and average coverage tell different stories.  An agent can retain most of its apparent competence while losing the reliability needed for an all-critical workflow.  This is especially important in code auditing, migration, access control, and safety checks, where one omitted field or preservation rule can invalidate the deliverable.

Sparse loss also explains why failures are difficult to notice.  The artifact is syntactically valid, most checks pass, tools were used, and the completion message sounds confident.  A user reviewing the result casually may see evidence of competence everywhere except at the one missing requirement.  Context rot is therefore not necessarily spectacular degradation; its practical signature can be a high-quality artifact with a low-probability critical hole.

\subsection{The Failure Chain Explains How Rot Becomes Silent}
The stage analysis suggests two central bottlenecks.  First, Acquisition is not enough: a rule can be present in the skill or even read from a file without becoming a durable task-specific obligation.  This Compilation gap produces missing arrays, forgotten cardinalities, and partial worklists.  Second, Verification is not independent when it is reconstructed by the same trajectory.  A self-check can test only what the agent still remembers and reproduce the generator's omission.

Completion amplifies both bottlenecks.  Once the agent's active representation no longer contains the missing requirement, its internal notion of ``done'' becomes easier to satisfy.  Contradictory evidence must explicitly reopen the task; otherwise a tool can expose the defect without affecting the final answer.  The mixed inventory's 38 silent closures among 44 failures makes this a practical concern, even though that sample cannot estimate a population rate.

This interpretation is stronger than a standalone taxonomy.  Lost requirements, editing drift, and failed checking are visible forms; the realization chain shows their relationship.  A requirement is first supplied, then selected, turned into work, acted on, and tested.  Rot can enter at any transition, and later stages often inherit rather than repair the loss.

\subsection{Why the Detailed Checklist Works}
The Detailed guard does not make the model generally more capable.  It changes the source of truth at the completion boundary.  Generic asks the current trajectory to regenerate its own validation set.  Detailed supplies a compact, externally frozen set of non-negotiable obligations.  If one requirement disappeared during Compilation, the final phase sees it again.

This result supports a simple design principle: keep rich background and evidence retrievable, but keep the pending critical obligations small, explicit, external to the growing transcript, and independently checkable.  The checklist and executable verifier serve different roles.  The checklist helps the agent avoid omissions; the verifier detects any remaining ones.  Both must be versioned with the skill because a stale external list creates a new failure source.

The evidence does not yet justify a complex new runtime.  For structured audit tasks, an ordinary checklist and deterministic script may capture most of the available benefit.  The broader Contract-Aware Skill Runtime proposed initially remains a research direction: compile natural-language instructions into explicit obligations, isolate stages, externalize state, and block completion on unresolved evidence.  Each mechanism should be tested separately before being presented as a system contribution.

\subsection{What Coding-Agent Scaffolds Can and Cannot Fix}
Coding agents can mitigate context load by searching the repository and bringing a small evidence subset into the active working set.  GPT-5.5's Bundled Direct and Codex results are consistent with this mechanism: both succeed, while Full Direct often fails.  This agrees with recent evidence that coding agents can process large repositories through selective traversal~\cite{cao2026longcontextagents}.

Selection addresses Acquisition load, not the full realization chain.  The main model finds the files yet still loses T002 during Compilation and repeats the omission during Verification.  Compaction can create a similar tradeoff: it reduces volume but may remove unresolved obligations or provenance.  Scaffolds should therefore externalize both the selected evidence and a small pending-requirement ledger.  Tool output that contradicts the expected result should automatically reopen the relevant item.

\subsection{Delegation Moves the Reliability Boundary}
A fresh sub-agent protects against inherited conversational clutter only at its start.  Once its own trajectory becomes long, the same obligation-survival problem reappears locally.  The opaque-contract run makes the distinction observable because random values cannot be reconstructed from semantics: after compaction, the worker either still has the contract, requests it again, fails, or obtains it from another source.

The exact contaminated artifact is also a warning about evaluation.  Final-state tests measure what was produced, not how missing state was recovered.  For tasks with evidence, privacy, licensing, or isolation constraints, success must combine artifact checks with enforceable read and write boundaries.  Prompt-only restrictions and self-disclosure are useful diagnostics, but an OS-level allowlist and external access trace are required before estimating violation rates.

Partial refresh further suggests that the durable unit is an executable contract, not just task data.  A refresh must include pending obligations, route, output path, serialization, evidence boundary, and completion condition.  Whether complete end refresh repairs the observed compaction loss remains open because that arm failed before receiving the intervention.

\subsection{Implications for Evaluation}
Context-rot evaluations should report at least four quantities together.  First, an onset curve or an honest bound on where degradation was and was not reproduced.  Second, repeated failure frequency with uncertainty.  Third, severity under both strict success and partial coverage.  Fourth, a propagation analysis that distinguishes retrieval, task formation, action, verification, and completion.  Omitting any view creates a misleadingly simple conclusion.

Clean capability gates and resistant tasks are equally important.  A clean-failing subject cannot identify context-induced loss, while a long-passing task falsifies universal claims.  Resultless attempts also need their own denominator.  The deployed system can fail because of context use or because the runner never returns an artifact; combining them corrupts diagnosis, while hiding retries overstates end-to-end reliability.

\subsection{Research Agenda}
The most useful next study is broader rather than deeper on this single task.  It should preregister at least five independently authored white-box audit skills, multiple context levels, ten or more valid runs per cell, a capped attempt policy, and at least two models.  Hierarchical analysis can then estimate task-to-task variation and a distribution of onset points.  Detailed checklists should be crossed with every task to determine whether restoration of external obligations generalizes beyond array and cardinality constraints.

The delegated-worker mechanism additionally needs a preregistered dose ladder and at least eight fresh repetitions of no refresh, sham refresh, data-only refresh, and complete-contract refresh.  Those arms should branch from the same post-load checkpoint, persist raw receipts outside conversational state, and enforce the read allowlist at the operating-system boundary.

Only after those replications should richer runtime mechanisms be added.  The immediate scientific target is not to prove that every skill eventually rots.  It is to identify which requirements, under which contexts and stages, lose enough reliability to matter.

\section{Threats to Validity}
\label{sec:threats}

\subsection{Construct Validity}
We observe behavior, not attention, memory, comprehension, or private reasoning.  A requirement absent from the artifact may have been internally represented and then lost later.  Stage labels therefore identify the earliest visible failure, not a neural mechanism.  Likewise, context rot is operationalized as a reliability change under controlled context manipulation; it is not evidence of irreversible model decay.

Character count is reproducible but only a rough size measure.  Relevant and Irrelevant are equal in characters, not necessarily tokens, semantic density, file layout, retrieval paths, or number of apparent tasks.  The 299K result applies to two concrete context constructions.  A future study should report tokenizer-specific lengths and manipulate structural factors independently.

The delegated-worker extension strengthens delivery evidence with per-call character counts and ordered canaries, but still reports characters rather than tokenizer-specific tokens.  Its ``model-visible'' label means that each tool result was returned intact to the worker interface; it does not prove which content survived later compaction or received internal attention.

The 24 checks are strongest for schema presence, types, cardinality, protected content, and task-specific discriminators.  They cannot prove the overall depth of an open-ended security audit.  ``Task success'' therefore means success under the frozen operational requirements, not proof that every possible vulnerability insight is correct.

The realization chain simplifies interacting failures.  A missing compiled obligation can cause a bad edit and an incomplete check; the summary assigns the earliest defensible visible stage.  Independent double-coding would improve reliability.  The broader 38/44 silent-completion and 2,857/2,916 coverage figures come from mixed historical inventories and are used only to show failure forms, never as population estimates.

\subsection{Internal and Conclusion Validity}
Agent output is stochastic, and the gateway may route a nominal model identifier to changing backends.  We freeze prompts, workspaces, settings, and evaluator revisions, but cannot freeze provider internals.  The main cells contain only ten valid runs.  Their Wilson intervals are wide, and Clean-versus-long Fisher tests yield $p=0.0698$.  We therefore emphasize observed effect sizes and replication need rather than statistical confirmation.

Sample sizes were extended after inspecting pilot results.  Irrelevant moved from 0/5 to 3/10 passes, showing the danger directly.  The final analysis is exploratory rather than preregistered.  The Detailed--Generic comparison reaches $p=0.0325$, but a single $n=10$ comparison may overestimate the benefit and does not establish generality.

Checks within an artifact are correlated.  We do not treat 240 check results as independent trials.  Coverage CRR is a descriptive severity measure; significance is computed only from run-level strict outcomes.  Sparse factor, cross-model, and scaffold probes receive no inferential ranking or multiple-comparison claims.

Infrastructure missingness is substantial and imbalanced.  Twenty-eight extension attempts are excluded, including 20 in Irrelevant.  If difficult generations disconnect more often, scoreable-run success may be optimistic; if disconnects are unrelated incidents, counting them as agent failures is unfair.  The logs cannot distinguish these cases.  We therefore expose both scoreable-run quality and attempt-level scoreability.

The decisive sub-agent result is one valid OLD trajectory and was discovered after earlier stress probes; it estimates neither a probability nor a threshold.  The codebook-refresh arm is a separate fresh trajectory, not a branch from an identical post-compaction checkpoint, and the full-contract arm fails its Acquisition gate.  Native compaction counts are system events, but the claim that the mapping was absent and the prohibited reads occurred depends on the worker's disclosure because the scaffold exposes no independent raw access trace.  We therefore classify the artifact as contaminated mechanism evidence, not as a silently observed violation rate.

\subsection{External Validity}
The strongest evidence concerns one production-derived structured code-audit task, one main model, and Codex.  The stable Revert SL-007 counterexample shows that even nearby tasks need not rot under the same gross size.  Four candidate tasks fail their clean gates, leaving no second independent clean-pass/long-fail replication.

Structured artifacts make requirement loss easier to detect than free-form code patches, browser tasks, or research reports.  Conversely, all-critical audit requirements may make sparse loss more consequential than benchmarks granting partial credit.  The qualitative propagation chain may transfer, but its frequency and stage distribution must be re-estimated in each domain.

Five model families are sampled unevenly, and gateway protocol support prevents a same-model Codex--Claude Code comparison.  Scaffold conclusions are consequently about selective working sets, not product superiority.  The Luna-equivalent cost schedule supports reproducibility but is not an actual provider bill.

The opaque mapping is intentionally unlike ordinary code semantics.  It isolates retention because a lost random value cannot be plausibly inferred, but may overstate the fragility of compressible instructions and understate the complexity of repository-scale evidence.  The negative reused-history and namespace controls also operate at only one modest dose each.  Delegation results must be repeated with semantic contracts, provenance tasks, and independently authored skills.

\subsection{Reproducibility and Researcher Judgment}
Early pilot workspaces contained answer-bearing artifacts.  The main experiment uses an allowlist sanitizer, and contaminated runs are excluded from effect claims.  Sanitization improves causal control but removes some realistic repository history.  In the OLD sub-agent arm, the raw 300-row receipt was not persisted before supervisor-context compaction; the preserved aggregate gate facts and verbatim disclosure remain available, but no rows were reconstructed.  The codebook-refresh and failed full-contract receipts are retained raw.  A public package should include the sanitizer, frozen requirement manifest, positive and negative checker fixtures, prompts, valid-run manifest, excluded-attempt audit, scoring scripts, delegated-worker receipts and access traces, and exact analysis commands while removing secrets and proprietary fragments.

\section{Related Work}
\label{sec:related}

\subsection{Effective Context and Context Rot}
Long-context research distinguishes nominal capacity from effective use.  Lost in the Middle shows strong positional effects~\cite{liu2023lostmiddle}; LongBench evaluates long-document understanding across tasks~\cite{bai2024longbench}; and RULER uses controlled probes to estimate usable context size~\cite{hsieh2024ruler}.  Chroma's controlled study varies length, distractor similarity, and structure and popularizes \emph{context rot} for performance degradation as input grows~\cite{hong2025contextrot}.  These works motivate our definition, but mostly evaluate retrieval, question answering, or bounded operations rather than a multi-stage coding workflow.

Recent work moves context rot closer to agentic settings.  Xia et al. study long-horizon search and observe premature uncertain answers as accumulated context grows, then evaluate context management and rejection sampling~\cite{xia2026searchrot}.  Martin and Roger show that long-transcript coding-agent monitors miss dangerous actions more often after extensive benign activity and that periodic reminders partially help~\cite{martin2026classifierrot}.  Our setting differs in actor and observability: the coding agent must realize a fixed skill, and we can trace each required artifact property from instruction to completion.  We additionally separate frequency, severity, onset, and lifecycle stage.

Software-engineering work also uses \emph{context rot} for stale persistent configuration such as \texttt{CLAUDE.md} or \texttt{AGENTS.md}~\cite{treude2026stalecontext}.  That consistency problem is complementary but distinct.  We hold instructions correct and repository state fixed, then study behavioral degradation caused by the execution context.

\subsection{Agentic Instruction Following and Skills}
AgentIF evaluates long, structured instructions with tool and conditional constraints and finds substantial difficulty with realistic constraint sets~\cite{qi2025agentif}.  SkillsBench treats skills as first-class artifacts across diverse tasks, comparing no-skill, curated-skill, and generated-skill conditions~\cite{li2026skillsbench}.  SkillLearnBench evaluates continual skill generation at skill, trajectory, and outcome levels~\cite{zhong2026skilllearnbench}.  Together, these studies show that procedural guidance can help but is not uniformly realized.

Our study fixes the skill rather than measuring skill uplift or skill generation.  It asks how the same mandatory requirements survive different contexts.  Requirement-level checks are not proposed as a novel monitoring language; they are the measurement instrument that reveals whether a high-level skill promise remains active in the final artifact.

\subsection{Agent and Coding-Agent Evaluation}
AgentBench covers multiple interactive environments~\cite{liu2023agentbench}; \taubench{} emphasizes final environment state and repeated-trial reliability~\cite{yao2024taubench}; and SWE-bench uses executable repository issues~\cite{jimenez2024swebench}.  SWE-agent and OpenHands show the importance of agent-computer interfaces, file tools, and persistent workspaces~\cite{yang2024sweagent,wang2024openhands}.  Cao et al. argue that coding agents can operate as long-context processors by traversing files instead of receiving a monolithic prompt~\cite{cao2026longcontextagents}.

These benchmarks make success concrete, but aggregate scores do not explain whether a fixed requirement was never acquired, lost during task formation, violated during editing, or ignored during validation.  Our white-box task sacrifices breadth to make that propagation observable.  The scaffold probe also tests the selective-working-set explanation directly, although unequal budgets prevent a product ranking.  The delegated-worker extension further treats artifact correctness and access provenance as separate outcomes, exposing recovery that a final-state benchmark would accept.

\subsection{Failure as a Trajectory}
Coding-agent failure studies increasingly analyze the path rather than only the last repository state.  Majgaonkar et al. show that failed agents can localize relevant files yet miss the required modification~\cite{majgaonkar2025codeagentbehaviour}.  Zhao et al. study onset, evolution, and recovery across thousands of CLI trajectories~\cite{zhao2026failureprocess}.  Tang et al. analyze more than 20,000 real sessions and identify user-visible misalignment including rule violations and inaccurate progress reports~\cite{tang2026failusers}.

Those studies are broader and much larger.  Our narrower contribution is causal structure at the task level: context is deliberately changed while the skill, artifact, and checks remain fixed.  The five-stage realization chain is not intended to replace general failure taxonomies.  It is a way to explain how context-associated loss travels from a known instruction to a silently invalid artifact.

\subsection{Position of This Work}
The closest intersection is effective-context evaluation plus agentic constraint following plus trajectory analysis.  This paper adds an in-depth white-box case in code auditing and organizes the evidence around six empirical questions: when rot appears, how often it appears, how much reliability is lost, how requirements disappear through execution, what helps, and whether delegation contains the failure.  Its novelty is this joint behavioral account and its boundary evidence, not the term \emph{context rot}, requirement monitoring itself, or a universal failure taxonomy.

\section{Conclusion}
\label{sec:conclusion}
This paper studies context rot as a behavioral reliability problem in agent skills.  Rather than asking only whether a long-context run passes, it asks when degradation appears, how often it occurs, how much capability is lost, where a mandatory instruction stops affecting a tool-using trajectory, what helps, and whether delegation contains the problem.

On the repeated Codex--\texttt{gpt-5.4-mini} audit task, failure rises from 2/10 in a 10,991-character Clean context to 7/10 in each 299,140-character long context.  Strict-success retention is 0.375, while requirement-coverage retention remains 0.933--0.949.  The result is therefore not general collapse but often sparse loss of critical obligations.  Scale pilots do not reveal a stable onset, and resistant tasks and models show that rot is conditional rather than universal.

The white-box traces place the main losses after retrieval: requirements disappear while being turned into active tasks, edits drift from local constraints, validation inherits the same blind spot, and completion closes over unresolved defects.  A detailed external checklist restores the obligation set and passes 10/10 runs, compared with 5/10 for generic self-checking.  Selective coding-agent retrieval can also reduce the working set, but does not protect every stage.

Delegation likewise moves rather than removes the reliability boundary.  In one fresh-worker run, a clean 2.4M-character load and three native compactions precede loss of an opaque contract; the worker reaches a byte-exact artifact only through disclosed prohibited reads.  This is mechanism evidence, not a 2.4M threshold.  It shows that artifact correctness, obligation survival, and access provenance are distinct success dimensions.

The present study has one repeated positive task, sparse boundary probes, and one decisive delegated-worker trajectory without OS-enforced read isolation, so it cannot estimate a universal prevalence or threshold.  The next step is a preregistered multi-task study with several context levels, fixed attempt caps, repeated checklist interventions, and replicated no-refresh versus complete-contract-refresh delegation arms.  The central lesson is already clear: context capacity should not be reported as if it were reliable procedural capacity.  For coding agents, context rot must be measured as a distribution over tasks and stages, with onset, frequency, severity, silent propagation, and evidence boundaries reported together.

\section*{Data Availability}
\label{sec:data-availability}
The research package contains the original proposal, requirements manifest, clean-workspace builders, condition prompts, runners, checker tests, frozen checker and summary scripts, valid-run manifests, per-run records, returned usage, failure catalog, interrupted-attempt audit, cost ledger, and delegated-worker fixtures and receipts.  Raw result directories retain failed and interrupted attempt suffixes rather than overwriting retries.  The package also records which early workspaces were contaminated by answer-bearing history and excludes them from the reported independent mitigation result.  For the delegated-worker study, raw receipts are preserved for the codebook-refresh and failed full-contract arms; the OLD raw 300-row receipt was lost before external persistence, so only its contemporaneous aggregate gate facts and verbatim disclosure are released and no ledger rows are reconstructed.  The anonymized artifact package is available from the author pending deposit in a permanent public archive; model credentials, internal gateway identifiers, proprietary source text, and user-specific absolute paths are excluded.

\appendix
\section{Failure-Coding Guide}
\label{app:coding-guide}

This appendix makes the descriptive classification reproducible.  Coding begins only after the frozen checker has assigned PASS or FAIL.  The coder reads the initial artifact, final artifact, tool log, and completion message.  Private reasoning is neither requested nor inferred.

\subsection{Ordered Decision Procedure}
For each failed run, apply the following rules in order:
\begin{enumerate}
  \item \textbf{Separate non-agent faults.}  If no usable artifact exists because the process never starts, the response disconnects before work, or the host terminates the runner, label the attempt Non-agent/Runtime.  If the saved artifact satisfies the intended requirement but the checker rejects semantically equivalent content, label the run Non-agent/Evaluator and correct the score transparently.
  \item \textbf{Test for a lost requirement.}  Compare the frozen requirement list with any visible plan, generated validator, and final artifact.  If a mandatory item has no satisfying output and no evidence that it remained active, label the first visible failure Lost Requirement.
  \item \textbf{Test for editing drift.}  If the requirement remains visible but the final edit violates it---for example, a protected field changes or a nearby problem replaces the requested one---label the failure Editing Drift.
  \item \textbf{Test for failed checking.}  If the artifact, a checker invocation, or tool output visibly contradicts the completion claim and the agent does not repair or reopen the task, label the failure Failed Checking.
  \item \textbf{Record completion separately.}  Mark \texttt{silent=yes} when the final response states success, safety, completion, or no unresolved work despite a critical failure.  Completion does not replace the first-visible class.
\end{enumerate}

The ordered rule prevents double-counting the same chain.  A missing array followed by an incomplete self-check is primarily Lost Requirement and secondarily Failed Checking.  A protected field explicitly listed in the task but overwritten during editing is Editing Drift.  A tool-output contradiction in an otherwise complete artifact is Failed Checking.

\subsection{Class Boundaries and Counterexamples}
\begin{table*}[t]
  \caption{Inclusion and exclusion examples for the four classes.}
  \centering
  \small
  \begin{tabularx}{\linewidth}{p{0.18\linewidth}p{0.36\linewidth}X}
    \toprule
    Class & Include & Do not include \\
    \midrule
    Lost requirement & Required array absent; full subtask untouched; cardinality rule absent from generated checks & Present rule implemented incorrectly for a value \\
    Editing drift & Protected subtree changed; schema type copied from adjacent field; nearby root cause substituted & Requirement never appears in work product \\
    Failed checking & Contradictory tool result ignored; validator omits decisive check; premature safe closure & Error exists but no checking evidence is visible \\
    Non-agent failure & First-token disconnect; host kill; verified semantic false negative & Valid artifact that fails a frozen requirement \\
    \bottomrule
  \end{tabularx}
\end{table*}

The classification is intentionally conservative about causality.  ``Lost requirement'' does not mean the model never read the text; it means the requirement is absent from the observable active work.  ``Failed checking'' does not mean the model internally understood the contradiction; it means the contradiction was available before the completion decision.

\section{Condition and Check Inventory}
\label{app:conditions}

\subsection{Main Check Families}
The 24 checks cover four audit tasks, T001--T004.  Each task contributes checks from the following families, with task-specific applicability:
\begin{itemize}
  \item required plural fields exist and use arrays, including empty arrays where permitted;
  \item field cardinalities remain within the stated maximum;
  \item task-specific evidence and discriminators are present;
  \item output values use the required schema types;
  \item protected identity and core fields equal their initial values; and
  \item the final artifact is parseable, structurally complete, and limited to the requested edit scope.
\end{itemize}

The exact requirement manifest and checker source are release artifacts.  We describe families rather than proprietary field semantics in the anonymous paper, but the run-level checker output names each failed predicate.

\subsection{Experimental Matrix}
\begin{table*}[t]
  \caption{Status of the experimental matrix.  ``Repeated'' denotes ten valid runs in each primary cell; other cells are descriptive probes.}
  \centering
  \small
  \begin{tabularx}{\linewidth}{p{0.22\linewidth}p{0.26\linewidth}p{0.16\linewidth}X}
    \toprule
    Experiment & Subject & Repetition & Role \\
    \midrule
    Main contexts & Stage~4CD, GPT-5.4 mini, Codex & 10 each & Clean/Relevant/Irrelevant comparison \\
    Generic/Detailed & Stage~4CD, GPT-5.4 mini, Codex & 10 each & Checklist intervention \\
    Revert SL-007 & GPT-5.4 mini, Codex & 2/3/3 & Stable counterexample \\
    Revert SL-005 & GPT-5.4 mini, Codex & 2/5/5 & Cross-task boundary \\
    Five-model probes & Stage~4CD & 1--3 cells & Model heterogeneity and clean gates \\
    Scaffold variants & Stage~4CD & 1--3 cells & Direct/Bundled/Full/Codex mechanism probe \\
    Context-factor pilots & Multiple workflows & 1--3 cells & Rule/history/position failure discovery only \\
    Delegated local context & Fresh/reused sub-agents & Mostly one run/arm & Compaction, refresh, provenance, and negative controls \\
    \bottomrule
  \end{tabularx}
\end{table*}

\subsection{Why Some Pilot Runs Are Not Pooled}
Pilot conditions differ in task, model, scaffold, prompt construction, contamination status, and verifier maturity.  Pooling them would create a large but uninterpretable sample.  The historical inventory is used only to show observed failure forms, silent closure, and the logical difference between check coverage and strict success.  Effect estimates use the sanitized repeated cells only.

\section{Interrupted-Attempt Audit}
\label{app:attempt-audit}

The extension preserves every excluded suffix.  Table~\ref{tab:excluded-types} lists the operational categories and their treatment.

\begin{table*}[t]
  \caption{Excluded attempts in the $n=10$ extension.}
  \label{tab:excluded-types}
  \centering
  \small
  \begin{tabularx}{\linewidth}{XrX}
    \toprule
    Type & Count & Scoring treatment \\
    \midrule
    Process could not start & 7 & No run directory or artifact; operational failure \\
    Stream disconnected before first token & 14 & Initialized root but no model work; operational failure \\
    Host SIGTERM/exit 143 or exit 137 & 4 & No complete scoreable record; operational failure \\
    Other resultless/indeterminate start & 2 & Retained as unresolved operational failure \\
    Null usage and no usable edit & 1 & Excluded from valid $n$; retained in audit \\
    \midrule
    Total & 28 & Never converted into model PASS or FAIL \\
    \bottomrule
  \end{tabularx}
\end{table*}

Transient reconnects inside runs that eventually return complete usage and checker output are not excluded.  A checker FAIL is always a valid model outcome.  This distinction prevents a retry from erasing a genuine artifact failure while also preventing an empty stream from being mislabeled as an audit mistake.

The recommended future policy is attempt-based: preregister a maximum number of attempts per condition, randomize condition order, report both scoreable-attempt rate and pass rate among scoreable attempts, and perform sensitivity analyses that bound end-to-end success if every missing attempt is treated as either a pass or a failure.  Such bounds are wide in the present study, which is why we avoid using the operational data to strengthen the context-effect claim.

\section{Usage and Cost Accounting}
\label{app:cost}

The cost ledger records logical tokens as the sum of uncached input, cached input, and output tokens returned by each backend.  We convert these counters with a single reference schedule rather than mixing provider-specific prices:
\begin{equation}
  \text{Cost}_{\mathrm{eq}} = 2I_u/10^6 + 0.2I_c/10^6 + 6O/10^6,
\end{equation}
where $I_u$ is uncached input, $I_c$ is cached input, and $O$ is output.  Applying the formula to 16,469,855 uncached input, 147,916,544 cached input, and 3,056,150 output tokens gives USD~56.022111.

The $n=10$ extension alone accounts for 32,921,773 logical tokens and USD~16.280576 equivalent cost across 29 new valid runs.  Attempts without returned usage are absent from monetary totals, so the cost is a confirmed lower bound.  The ledger should not be read as the price of reproducing the paper on another gateway because caching semantics and retry rates may differ.

\section{Reproduction Checklist}
\label{app:reproduction}

A reproduction should complete the following steps in order:
\begin{enumerate}
  \item pin the skill package, task fixture, runner revision, model identifier, reasoning effort, and checker revision;
  \item build the workspace from the sanitizer allowlist and verify that answer-bearing files are absent;
  \item run positive and targeted negative checker fixtures before any model calls;
  \item predeclare conditions, valid-run criteria, maximum attempts, retry policy, sample size, and statistical comparisons;
  \item randomize or interleave condition order to reduce temporal gateway confounding;
  \item save the exact prompt, starting artifact hash, changed files, raw output, tool log, usage, process status, and checker result for every attempt;
  \item for delegated runs, persist per-call visibility receipts and access traces outside both worker and supervisor conversational state, and enforce read/write allowlists technically;
  \item retain failed and resultless attempts under unique identifiers rather than overwriting them;
  \item compute strict success, requirement coverage, failure-rate Wilson intervals, and planned Fisher tests from a frozen valid-run manifest;
  \item classify failures only after scoring, using the ordered guide in Appendix~\ref{app:coding-guide}; and
  \item publish an anonymized package with analysis commands, exclusion audit, cost ledger, and a record of any evaluator rescores.
\end{enumerate}

\section{Additional Research Questions for a Larger Study}
\label{app:future-rqs}

The current results motivate six focused extensions.  First, does failure probability change monotonically with context length when all other dimensions are fixed?  The present study compares only one clean and one long endpoint.  Second, does the Detailed checklist replicate across independently authored audit skills, or is its benefit specific to array and cardinality constraints?  Third, which manipulation drives lost requirements: semantic similarity, number of active obligations, evidence position, stale history, or tool-output volume?  Fourth, can the same model be compared across Direct, Codex, and Claude Code under equal tool and token budgets?  Fifth, do the four visible classes remain sufficient when the output is a code patch or vulnerability finding rather than a structured task file?  Sixth, after the same verified post-load checkpoint, do sham, data-only, and complete executable-contract refreshes differ in their ability to prevent delegated-worker loss?

A preregistered study should prioritize task replication over deeper analysis of the current workflow.  A reasonable first target is at least five white-box audit skills, two models, three context levels, and ten or more valid runs per cell, with an attempt cap and hierarchical analysis across tasks.  Such a design would test whether the observed lifecycle pattern and checklist benefit generalize while estimating task-to-task variation rather than hiding it.

\begingroup
\small
\bibliographystyle{plain}
\bibliography{references}
\endgroup
\end{document}